# HEAT TRANSFER IN THE FLOW OF A COLD, TWO-DIMENSIONAL VERTICAL LIQUID JET AGAINST A HOT, HORIZONTAL PLATE

Jian-Jun SHU and Graham WILKS

*School of Mechanical & Aerospace Engineering, Nanyang Technological University, 50 Nanyang Avenue, Singapore 639798, mjjshu@ntu.edu.sg, http://www.ntu.edu.sg/home/mjjshu*

Abstract: A cold, thin film of liquid impinging on an isothermal hot, horizontal surface has been investigated. An approximate solution for the velocity and temperature distributions in the flow along the horizontal surface is developed, which exploits the hydrodynamic similarity solution for thin film flow. The approximate solution may provide a valuable basis for assessing flow and heat transfer in more complex settings.

2000 AMS Subjects Classification: 80A20

1. INTRODUCTION

In response to the abundance of practical applications the heat transfer associated with impinging jets has been the subject of numerous theoretical and experimental research studies reported in the literature [1-7]. This is particularly true in the context of condensers, as used in power generation. Here the draining fluid is accumulated condensate.

The problem to be examined concerns the film cooling, which occurs when a cold vertically draining sheet strikes a hot horizontal plate. Although a sheet of fluid draining under gravity will accelerate and thin, at impact it is reasonable to model the associated volume flow as a jet of uniform velocity $U_0$ and semi-thickness $H_0$, as is illustrated in Figure. The notation $Q = U_0 H_0$ is introduced for the flow rate and a film Reynolds number may be defined as $R_e = \dfrac{\rho Q}{\mu}$, where $\mu$ is the dynamic viscosity of the fluid.

The temperature condition within which heat transfer estimates will be obtained assumes a constant temperature $T_w$ at the plane and zero heat flux at the free surface. If the rate of viscous diffusion exceeds that of temperature diffusion, the point at which viscous effects penetrate the free surface will occur before the point at which the free surface first experiences the presence of the hot plane. This physical appraisal of the developing flow field provides the framework for the initial approximate method of solution. Schematically the flow may be represented as in Figure and divided into the following regions [8-10].

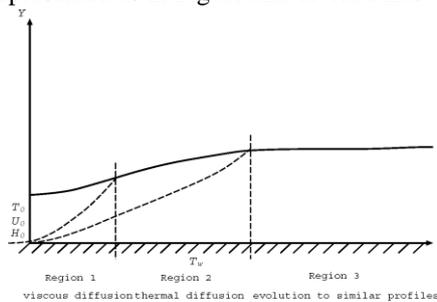

Figure: Basis of approximate solution

*Region* 1: The impinging jet essentially experiences an inviscid symmetric division and deflection. A viscous boundary layer develops against the horizontal plate within the deflected jet and eventually penetrates the free surface marking the end. A thermal boundary layer develops simultaneously, but for the Prandtl numbers greater than unity this will still be evolving at the end.

*Region* 2: A judicious choice of approximating profiles in Region 1 is designed to approximate immediate transition to the film similarity solution at the onset of Region 2. Consequently Region 2 is examined under the assumption that the full hydrodynamic similarity solution is applicable. The adjustment of the temperature field as thermal effects develop and penetrate the free surface within this hydrodynamic setting is monitored. The end is notionally reached when the presence of the hot wall is first detected at the free surface.

*Region* 3: In the film cooling setting, when there is zero heat flux at the free surface, the film will eventually reach a uniform temperature distribution, coinciding with the temperature of the wall. Again within the established hydrodynamics, Region 3 covers the evolution towards this asymptotic state once wall temperature effects penetrate the free surface.

## 2. GOVERNING EQUATIONS

The flow under investigation has been modeled as a steady, two-dimensional flow of incompressible fluid. On the assumption that the film thickness remains thin relative to a characteristic horizontal dimension, a boundary layer treatment of the equations leads to significant simplification. The following non-dimensional variables are introduced $x = \dfrac{X}{R_e H_0}$, $y = \dfrac{Y}{H_0}$, $\bar{h}(x) = \dfrac{H(X)}{H_0}$, $\bar{U} = \dfrac{U}{U_0}$, $\bar{V} = \dfrac{R_e V}{U_0}$, $\bar{\phi} = \dfrac{T - T_w}{T_0 - T_w}$. In the limit $R_e \to +\infty$, with $x$ remaining $O(1)$, the following equations are obtained

$$\frac{\partial \bar{U}}{\partial x} + \frac{\partial \bar{V}}{\partial y} = 0, \quad \bar{U}\frac{\partial \bar{U}}{\partial x} + \bar{V}\frac{\partial \bar{U}}{\partial y} = \frac{\partial^2 \bar{U}}{\partial y^2}, \quad P_r\left(\bar{U}\frac{\partial \bar{\phi}}{\partial x} + \bar{V}\frac{\partial \bar{\phi}}{\partial y}\right) = \frac{\partial^2 \bar{\phi}}{\partial y^2} \qquad (1)$$

where $P_r = \dfrac{C_p \mu}{\kappa}$ is the Prandtl number. The boundary conditions now read $\bar{U} = \bar{V} = 0$, $\bar{\phi} = 0$ on $y = 0$, $x \geq 0$; $\dfrac{\partial \bar{U}}{\partial y} = 0$, $\dfrac{\partial \bar{\phi}}{\partial y} = 0$ at $y = \bar{h}(x)$, $x \geq 0$; $\int_0^{\bar{h}(x)} \bar{U}\,dy = 1$ for $x \geq 0$. These have been quoted in the context of the fully developed film flow field, which is approached in Region 3. These solutions provide the basis for developing comprehensive approximate solutions for the complete flow field downstream of the symmetry point of impingement incorporating Regions 1, 2 and 3.

## 3. APPROXIMATE SOLUTIONS

*Region* 1: The equations governing the viscous and thermal boundary layers are exactly the same as (1), but the boundary conditions now read $\bar{U} = 0$, $\bar{V} = 0$, $\bar{\phi} = 0$ on $y = 0$, $x \geq 0$; $\bar{U} \to 1$, $\bar{\phi} \to 1$ as $y$ approaches the outer limits of the viscous and thermal boundary layers respectively $\bar{U} = 1$, $\bar{\phi} = 1$ at $x = 0$, $y > 0$. Their solutions for $P_r > 1$ indicate that the length scale of thermal diffusion can be significantly less than that of viscous diffusion. Viscous effects, in due course, must penetrate the free surface and the transition region of Figure is essentially a region of adjustment from the Blasius profile to the similarity profile. As the profiles are not greatly dissimilar, a device that in effect compresses the transition region to a single point is introduced. An approximate velocity profile

$$\bar{U}(x, y) = \bar{U}_s(x) f'(\eta), \quad \eta = \frac{y}{\delta(x)} \qquad (2)$$

is assumed, where $\delta(x)$ is the non-dimensional boundary layer thickness. A polynomial approximation to the velocity profile is more convenient. To maintain the aggregate and matching properties of $f'(\eta)$, and simultaneously exploit the convenience of a polynomial representation, a fourth-order polynomial approximation to $f'(\eta)$ has been obtained as $f'(\eta) = c\eta + (4 - 3c)\eta^3 + (2c - 3)\eta^4$. The profile is then used in a Kármán-Pohlhausen method of solution. Over Region 1 unretarded fluid is present when $x < x_0$, say where $x_0$ marks the point of penetration of viscous effects at the free surface, so that $\bar{U}_s(x) = 1$ and $\delta(x) < \bar{h}(x)$ over $0 < x < x_0$. For $x > x_0$ into Region 2 $\delta(x) \equiv \bar{h}(x)$ and $\bar{U}_s(x) < 1$ in a manner which, using the conservation of flow constraint, can be matched directly onto the asymptotic similarity solutions. The momentum integral equation reads $\dfrac{d}{dx}\int_0^{\delta(x)} \bar{U}(1 - \bar{U})\,dy = \left(\dfrac{\partial \bar{U}}{\partial y}\right)_{y=0}$ and using (2) leads to the solution

$\delta^2 = 19.775x$, where $\delta(x)=0$ has been assumed at $x=0$, which is valid in the limit of the underlying assumption. Invoking the conservation of volume flow at $x_0$, the end point of Region 1 leads to $\int_0^{\delta(x)} \overline{U}\,dy + (\overline{h}-\delta) = 1$, whence $\overline{h}(x) = 1 + \frac{3(4-c)\delta}{20}$. Since $\delta(x_0) = \overline{h}$, $x_0 = 0.136$ and matching the free surface velocity at $x = x_0$ leads to $l = 0.76$. The polynomial $f'(\eta)$ is consequently used in the subsequent developments of velocity and temperature distributions. It remains to establish the temperature characteristics in Region 1. The energy integral equation becomes

$$\frac{d}{dx}\int_0^{\delta_T(x)} \overline{U}(1-\overline{\phi})\,dy = \frac{1}{P_r}\left(\frac{\partial \overline{\phi}}{\partial y}\right)_{y=0}, \quad (3)$$

where $\delta_T(x)$ denotes the outer limits of the region of thermal diffusion. For $P_r > 1$, $\delta_T(x) < \delta(x)$ over $0 < x < x_0$. The notation $\eta_T = \frac{y}{\delta_T(x)}$ is introduced and the ratio $\frac{\delta_T}{\delta}$ is denoted by $\Delta$ so that $\eta = \Delta \eta_T$. The solution for $\delta_T(x)$ is again developed by assuming profiles for $\overline{U}$ and $\overline{\phi}$ as $\overline{U}(\eta) = f'(\eta)$, $\overline{\phi}(\eta_T) = f'(\eta_T)$, which ensures identical velocity and temperature distributions for $P_r = 1$ when also $\Delta = 1$. Assuming a constant ratio $\Delta$ leads to

$$P_r \Delta^2 = \frac{0.142}{D(\Delta)}, \quad (4)$$

where $D(\Delta) = \Delta(0.149 - 0.005\Delta^2 - 0.003\Delta^3)$. The values of $\Delta$ can obtained numerically for various Prandtl numbers. Notice that as a result of the choice of approximating profile the velocity distribution at the end of Region 1 exactly matches that of Region 2.

*Region* 2: The hydrodynamics are governed by the similarity solution where thermal diffusion continues to progress across the film. Accordingly the velocity at the free surface is no longer uniform, but is prescribed in non-dimensional terms. The film thickness $\overline{h}(x)$ and the viscous boundary layer thickness $\delta(x)$ now coincide as $\delta(x) = \overline{h}(x) = \frac{\pi}{\sqrt{3}}(x+l)$. The energy integral equation (3) remains appropriate. The presence of the free surface limits further viscous penetration and $\delta_T(x) \to \delta(x) = \overline{h}(x)$. In prescribing profiles $\eta_T = \frac{y}{\delta_T(x)}$ may again be utilized, but now $\Delta(x) = \frac{\delta_T(x)}{\delta(x)}$ is no longer constant, and must in fact tend to $1$ at the end. The following profiles are introduced into the energy equation: $\overline{U}(x,\eta) = \overline{U}_s(x)f'(\eta)$, $\overline{\phi}(x,\eta_T) = f'(\eta_T)$. The equation for $\delta_T(x)$ is accordingly

$\delta_T(x)\frac{d}{dx}\left\{\overline{U}_s(x)\delta_T(x)\int_0^1 f'(\eta)[1-f'(\eta_T)]d\eta_T\right\} = \frac{c}{P_r}$. This resultant first-order equation in $\Delta^2$ may now be integrated with initial data $\Delta(x_0;P_r)$ as far as $\Delta(x_1(P_r);P_r) = 1$ to give

$$\Delta^2(0.299 - 0.019\Delta^2 - 0.014\Delta^3) = \frac{0.476}{P_r}\ln\frac{x+l}{x_1+l} + 0.266. \quad (5)$$

$x_1$ marks the end, as predicted, using the polynomial profile. Beyond $x_1$ viscous and thermal effects are present throughout the film.

*Region* 3: To accommodate the adjustment of the film temperature to $T_w$, the following profiles are adopted $\overline{U}(x,\eta) = \overline{U}_s(x)f'(\eta)$, $\overline{\phi}(x,\eta) = \beta(x)f'(\eta)$, where now $\eta = \frac{y}{\overline{h}(x)}$. The energy integral equation now reads $\frac{d}{dx}\int_0^{\overline{h}(x)} \overline{U}(\beta - \overline{\phi})\,dy - \int_0^{\overline{h}(x)} \overline{U}\frac{d\beta}{dx}\,dy = \frac{1}{P_r}\left(\frac{\partial \overline{\phi}}{\partial y}\right)_{y=0}$. The result is an equation for $\beta(x)$ within the framework of prescribed film thickness, namely $0.142\frac{d}{dx}(\overline{U}_s\overline{h}\beta) - 0.61\overline{U}_s\overline{h}\frac{d\beta}{dx} = \frac{1.402}{P_r}\frac{\beta}{\overline{h}}$ and hence, $\beta(x) = \left(\frac{x_l + l}{x + l}\right)^{\frac{1.015}{P_r}}$, which satisfies the requirements $\beta(x_l(P_r)) = 1$ and has $\beta(x) \to 0$ at rates dependent on $P_r$.

## 4. RESULTS AND DISCUSSION

The approximate solution scheme outlined provides the comprehensive details of the flow and heat transfer characteristics for the model flow. The estimates of film thickness, velocity and temperature distributions, skin friction and heat transfer coefficients over the entire region downstream of the point of impingement can be obtained.

The elements of interest in engineering practice are the shear stress at the solid boundary, *i.e.* the skin friction and the rate of heat transfer at the boundary. The non-dimensional skin friction coefficient is given by $\overline{\tau} = \left(\frac{\partial \overline{U}}{\partial y}\right)_{y=0} = \begin{cases} 0.315 x^{-\frac{1}{2}} & \text{in Region 1} \\ 0.693(x+l)^{-2} & \text{in Regions 2 and 3} \end{cases}$. The integrable square root singularity is consistent with the Blasius boundary layer equivalent. The most significant film cooling design factor is the heat transfer across the film. The Nusselt number is given by

$$N_u = \left(\frac{\partial \overline{\phi}}{\partial y}\right)_{y=0} = \begin{cases} \frac{1}{\Delta(P_r)}\frac{0.315}{\sqrt{x}} & \text{in Region 1} \\ \frac{1}{\Delta(x;P_r)}\frac{0.773}{x+l} & \text{in Region 2} \\ \frac{0.773}{x+l}\left(\frac{x_l+l}{x+l}\right)^{\frac{1.015}{P_r}} & \text{in Region 3} \end{cases}$$

. The values of $\Delta(P_r)$ have been obtained from (4) and $\Delta(x; P_r)$ is the solution of (5).

## 5. CONCLUSIONS

An approximate and the elements of engineering practice, namely the skin friction and heat transfer coefficients for the flow of a cold two-dimensional jet against a hot, horizontal plate have been presented. Although at this stage a comparison between theory and experiment is unavailable, the work provides the basis for re-assessing condensation.